\documentclass{iaus}
\usepackage{graphicx}

\newcommand\sun{\odot}%
\newcommand{\kms}{km~s$^{-1}$}
\newcommand{\um}{$\mu$m}
\newcommand{\lohmax}{$L_{OH}^{max}$}

\title[Searching for OHMs at $z>1$] 
{Searching for new OH~megamasers out to redshifts $z>1$}

\author[K.W. Willett]   
{Kyle W. Willett$^{1,2}$}

\affiliation{$^1$Center for Astrophysics and Space Astronomy, University of Colorado\\ 
UCB 391, Boulder, CO 80304, United States \\
[\affilskip]
$^2$School of Physics and Astronomy, University of Minnesota \\
116 Church St. SE, Minneapolis, MN 55455, United States
\\email: {\tt willett@physics.umn.edu}}

\pubyear{2012}
\volume{287}  
\pagerange{}
\jname{Cosmic Masers from OH to H$_0$}
\editors{R. Booth, E. Humphreys \& W. Vlemmings, eds.}
\begin{document}

\maketitle

\begin{abstract}
We have carried out a search for 18-cm OH megamaser (OHM) emission with the Green Bank Telescope. The targeted galaxies comprise a sample of 121 ULIRGs at $0.09<z<1.5$, making this the first large, systematic search for OHMs at $z>0.25$. Nine new detections of OHMs are reported, all at redshifts $z<0.25$. For the remainder of the galaxies, observations constrain the upper limit on OH emission; this rules out OHMs of moderate brightness ($L_{OH}>10^3~L\sun$) for 26\% of the sample, and extremely bright OHM emission ($L_{OH}>10^4~L\sun$) for 73\% of the sample. Losses from RFI result in the OHM detection fraction being significantly lower than expected for galaxies with $L_{IR}>10^{12}~L_\sun$. The new OHM detections are used to calculate an updated OH luminosity function, with $\Phi\propto~L_{OH}^{-0.66}$; this slope is in agreement with previous results. Non-detections of OHMs in the COSMOS field constrain the predicted sky density of OHMs; the results are consistent with a galaxy merger rate evolving as $(1+z)^m$, where $m\leq6$. 
\end{abstract}

\firstsection 
\section{Introduction}

OH megamasers (OHMs) trace of some of the most extreme physical conditions in the universe - in particular, the presence of an OHM signals specific stages in the merger process of gas-rich galaxies. OHMs can thus be used as probes of their environments, both directly and indirectly. Characteristics of the maser emission itself can be used to measure extragalactic magnetic fields (via Zeeman splitting) and gas kinematics, while the {\it presence} of an OHM is a signpost for phenomena associated with galaxy mergers, including extreme star formation and merging black holes. OHMs are a unique tool in this respect due to their extreme luminosities and ability to be seen at cosmic distances. 

The total number of OHMs detected to date is still low. As of 2012, there are $\sim113$ OHMs published in the literature, with roughly 50\% discovered in the Arecibo survey of \cite[Darling \& Giovanelli (2000, 2001, 2002a)]{dar00,dar01,dar02}. No OHMs at a distance of greater than 1300~Mpc ($z=0.265$) have been detected, and no large, systematic searches for high-$z$ OHMs have been carried out. The association of OHMs with IR-bright merging galaxies, however, means that the density of OHMs is expected to be much higher at $z\simeq1-2$, coinciding with an increase in both merging rate and cosmic star formation. We have conducted a search for high-redshift OHMs using the Green Bank Telescope (GBT). 

\section{Sample selection and observations}

We constructed three samples of galaxies to search for high-redshift OHMs. None of the samples are fully complete or flux-limited, but draw on the catalogs of IR-luminous galaxies with well-defined redshifts that were available at the time.  

\underline{\it{IRAS PSCz galaxies not visible from Arecibo:}} The first sample of galaxies consisted of IRAS sources included in the redshift catalog of the PSCz survey. Galaxies were selected according to similar criteria as in the flux-limited Arecibo sample, but included objects lying outside the declination limits of Arecibo ($-1^\circ<\delta<38^\circ$). Galaxies selected for GBT observations had: a declination range of $-40^\circ<\delta<0^\circ$ or $\delta>37^\circ$, a redshift range of $0.10<z<0.25$, and a lower-luminosity threshold of $L_{60\mu m} > 10^{11.4} L_\sun$. 153 galaxies in the PSCz met these criteria, of which 47 of the brightest candidates were observed according to the LST windows during the early commissioning phase of the GBT in 2002.

\underline{\it{Sub-mm and ULIRG galaxies from the field:}} The second sample of potential OHM hosts was assembled from flux-limited catalogs of ULIRGs at higher redshifts. We began with 35 galaxies in the FSC-FIRST catalog \cite{sta00}, which consists of targets detected in both the IRAS Faint Source Catalog and the 20-cm VLA FIRST survey. This was supplemented with 5 IR-bright galaxy pairs and 26 sub-millimetre galaxies. All galaxies have $L_{IR}>10^{11}L_\odot$, with more than half having $L_{IR}>10^{12}L_\odot$. The highest redshift in this sample is at $z=1.55$. 

\underline{\it{Starburst galaxies from COSMOS:}} The third group of 19 OHM candidates was the last observed in our program, and made explicit use of the results from the first two samples. Targets were selected from the COSMOS field, a 2-deg$^2$ survey with deep spectral coverage from X-ray through radio wavelengths. Recent infrared \cite{wil11} studies show that while OHMs are found in infrared-bright galaxies, the OHM fraction is much higher for starburst-dominated galaxies vs. AGN. Selection of OHM candidates began with COSMOS galaxies detected by {\it Spitzer} at 70~\um, and then eliminating all targets except the LIRGs, ULIRGs, and HyLIRGs identified by \cite{kar10}. We removed all galaxies identified as AGN or with $L_{IR}<10^{12}L_\odot$. Finally, we culled the target list based on the expected RFI conditions near the observed frequency bands. By limiting the observed OH frequencies to cleaner regions, we have a broader margin for error on the galaxy redshift. The two windows used are at $\nu_{obs}=825-830$~MHz and $960-1005$~MHz, equivalent to redshifted OH at $0.97<z<1.01$ and $0.67<z<0.73$. 

{\bf Observations: } We observed the OHM candidates in several sessions at the GBT from 2002--2010, totaling approximately 150 hours. The majority of observations used the maximum available bandwidth of 50~MHz and 8192 channels. Integration times were selected with the goal of achieving $\sim1$~mJy rms per channel for each galaxy. The data were reduced using standard routines in GBTIDL, including extensive flagging for RFI. After flagging, we fit the radio continuum around the expected line center with a polynomial function of order $n=5$. This removed both intrinsic continuum structure from the target itself and any baseline structure not removed by the position-switching technique. After stacking, the spectra were smoothed to a rest-frame velocity resolution of 10~\kms. 

\begin{table}
  \begin{center}
  \caption{Properties of new OHM detections from the GBT survey}
  \label{tab1}
 {\scriptsize
  \begin{tabular}{|l|l|r|c|c|c|}\hline 
{\bf Galaxy} & {\bf $z_{hel}$} & {\bf $S_{1667}^{peak}$} & {\bf $W_{1667}$} & {log~$L_{OH}$} & {log~$L_{OH}^{pred}$} \\ 
             &  & {[mJy]} & {[MHz]} & {[$L_\sun$]} & {[$L_\sun$]} \\ \hline
IRAS 00461$-$0728 & 0.2427   & 8.93   & 0.937 & 3.58 & 2.94$-$3.30 \\
IRAS 01298$-$0744 & 0.136181 & 118.28 & 0.785 & 3.78 & 3.10        \\
IRAS 01355$-$1814 & 0.192    & 6.99   & 0.344 & 2.81 & 3.25        \\
IRAS 01569$-$2939 & 0.1402   & 40.67  & 1.821 & 3.93 & 2.95        \\
IRAS 10597$+$5926 & 0.196    & 17.56  & 0.915 & 3.85 & 3.13        \\
IRAS 12071$-$0444 & 0.128355 & 4.27   & 0.598 & 2.45 & 3.07        \\
IRAS 16090$-$0139 & 0.13358  & 9.00   & 3.110 & 3.53 & 3.35        \\
FF 0758+2851      & 0.126    & 4.67   & 0.396 & 2.67 & 2.39        \\
FF 2216+0058      & 0.212    & 18.40   & 0.347 & 3.53 & 2.58        \\ \hline
  \end{tabular}
  }
 \end{center}
\vspace{1mm}
\end{table}

\section{Survey results}

\begin{figure}[t]
\begin{center}
 \includegraphics[width=5.4in]{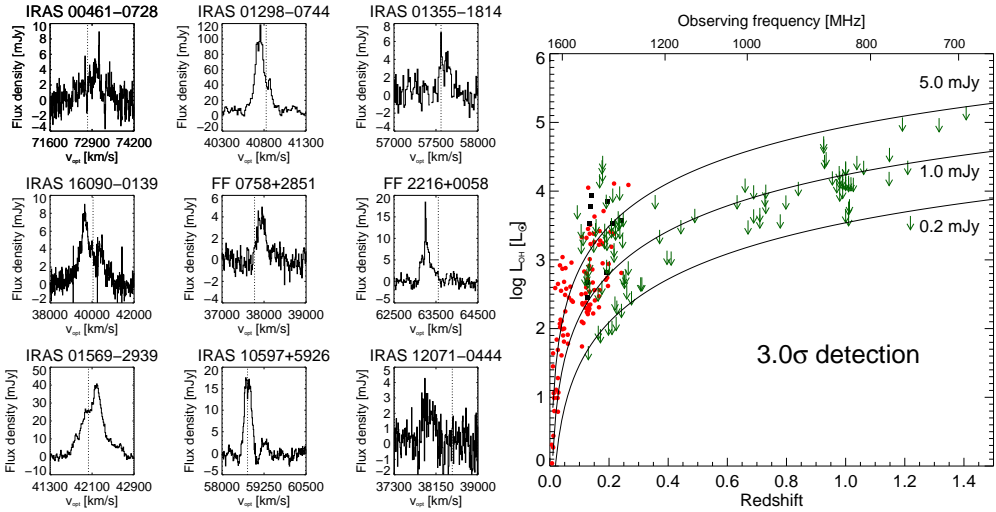} 
 \caption{Left: 18-cm spectra of the nine new OHMs discovered with the Green Bank Telescope. Right: Integrated OH luminosities all known OHMs, including data from the literature (red) and the new GBT detections (black). Upper limits from the GBT survey are indicated with the green arrows. The black curves show detection thresholds for a $\Delta$$v=150$~\kms~OHM at a range of telescope rms sensitivities as a function of redshift. }
   \label{fig1}
\end{center}
\end{figure}

\underline{OH megamasers:} Out of 128 galaxies observed for OH, we detected new OH megamaser emission in nine objects (Figure~\ref{fig1}). Seven of the detections were PSCz galaxies from the first sample, while the other two were from the FSC-FIRST catalog in the second sample. All nine OHMs have redshifts near the lower end of the sample distribution, with the most distant lying at a redshift of $z=0.2427$. The observed frequencies are in the range of $1300-1500$~MHz, which is covered by the L-band receiver and has relatively little RFI compared to the GBT prime focus bands. The OH emission has been confirmed with GBT follow-up observations for five of the galaxies. We also confirmed the detection of the previously-discovered OHM IRAS~09539+0857 \cite{dar01}.

Table~\ref{tab1} lists the 18-cm radio properties of the new OHM detections. We give the galaxy's optical redshift ($z_{hel}$), peak flux density of the OHM ($S_{1667}^{peak}$), the ratio of the integrated 1667~MHz emission to its peak flux density ($W_{1667}$), measured OH luminosity (log~$L_{OH}$), and predicted OH luminosity ($L_{OH}^{pred}$) based on the $L_{OH}-L_{FIR}$ relationship in \cite{dar02}.

\underline{OH non-detections:} 112 galaxies showed no confirmed detections of OH. The \lohmax~for each galaxy was (conservatively) derived assuming a boxcar line profile with a linewidth $\Delta v=150$~\kms~and a 1.5$\sigma$ detection. The rms was measured from baseline-subtracted continuum centered on the optical redshift of the galaxy and in a frequency range sufficient to cover the uncertainty in the optical redshift ($\Delta\nu_{obs}=\Delta z\times\nu_{rest}/[1+z]$).  

\section{Updating the OH luminosity function}

One of the goals of performing a search for OHMs at higher redshifts was to improve the measurements of the OH megamaser luminosity function (LF). \cite{dar02a} used the results of the flux-limited Arecibo survey to construct a well-sampled LF between $10^{2.2}~L_\sun<L_{OH}<10^{3.8}L_\sun$, which followed a power law in integrated line luminosity of $\Phi[L]\propto~L_{OH}^{-0.64}$~Mpc$^{-3}$~dex$^{-1}$. This measurement was limited to a narrow redshift range, spanning $0.1<z<0.23$. We constructed a new OH LF by combining the GBT and Arecibo OHM detections, using the $1/V_a$ method and combining limits on both spectral line and continuum emission. We fit a power-law to all bins with more than one detection, yielding:

\begin{equation}
\label{eqn-ohlf_combined}
{\rm log}~\Phi[L]=(-0.66\pm0.14)~{\rm log}~L_{OH} - (4.91\pm0.41),
\end{equation}

\noindent where $\Phi$ is measured in Mpc$^{-3}$~dex$^{-1}$ and $L_{OH}$ in $L_\sun$. The new detections only change the slope measured by \cite{dar02a} by $-0.02$ and the offset by $+0.10$. Both values are well within the uncertainties of the combined LF, as well as that of the original Arecibo LF. 

Assuming the Malmquist-corrected relationship of $L_{OH}\propto L_{IR}^{1.2}$ from \cite{dar02}, this gives $\Phi[L_{IR}]\propto L_{IR}^{-0.83\pm0.18}$. We compare this to the LF of ULIRGs in the local Universe using the AKARI measurements of \cite{got11}. Folding in the OHM fraction derived from the combined Arecibo and GBT samples results in a slope of $(-0.6\pm0.2)$, a much shallower value than that measured from the AKARI galaxies $(-2.6\pm0.1)$. This inconsistency may suggest either that OHMs are highly saturated or that the maser strength is only weakly correlated with global properties such as $L_{IR}$ \cite[(Darling \& Giovanelli 2002a)]{dar02}. The OHM LF of \cite{bri98} assumed a quadratic OH-IR relation corresponding to unsaturated masing; a decrease in saturation could potentially steepen the OHM LF up to a slope of $-1.5$. 

\section{Constraining the evolution of the cosmic merger rate}

One of the ultimate goals of high-redshift OHM surveys is to use megamasers as tracers of the populations of merging galaxies as a function of redshift. Models of the merger rate as a function of redshift are typically parameterized with an evolutionary factor of $(1+z)^m$. The value of $m$, however, is not well-constrained \cite[({\it e.g.,} Kim \& Sanders 1998; Bridge \etal~2010)]{kim98,bri10}. Sufficiently deep surveys of OHMs can provide an independent measurement of the parametrization of the merging rate. We calculated the predicted sky density of OHMs as a function of redshift:

\begin{equation}
\label{eqn-ohm_skydensity}
\frac{dN}{d\Omega d\nu}[z] = \frac{c D_L^2}{H_0\nu_0\sqrt{(1+z)^3\Omega_M + \Omega_\Lambda}} \left(\frac{b}{a~{\rm ln} 10}\right) \times \nonumber \\
\left((L_{OH,max})^a - (L_{OH,min})^a\right),
\end{equation}

\noindent where $a$ and $b$ are parameters of the OHM LF from Equation~\ref{eqn-ohlf_combined} ($\Phi[L_{OH}]=b L_{OH}^a$), $L_{OH,min}$ is the minimum OH luminosity that can be observed at a given sensitivity level, and $L_{OH,max}$ is the upper physical limit on OHM luminosity \cite[(Darling \& Giovanelli 2002b)]{dar02a}. 

Using the upper limit of zero OHMs detected in the COSMOS field, we place an upper limit on the merger rate of $m\lesssim6$. While this is still within uncertainties for the highest estimated values of $m$, the COSMOS limit is an important first step in using OHMs as an independent tracer. Further measurements of OH deep fields at higher redshifts will be crucial for a more accurate constraint. 

\section{Future OHM searches}

The final results for the GBT OHM survey yielded a much lower detection rate (9/121 = 7\%) of megamasers than expected. Based on the high detection fraction for galaxies with $L_{IR}>10^{12}~L_\sun$ from the Arecibo survey, in addition to our careful selection of starburst-dominated galaxies, we had predicted a success rate of 20--30\%. We attribute one of the primary causes of the low OHM fraction to be the sensitivity of the observations. Figure~\ref{fig1} shows the upper limits for OHM candidates from our survey, along with the necessary rms sensitivity to detect OH lines as a function of redshift. To detect median-luminosity OHMs at $z=1$ will require rms levels of 100~$\mu$Jy, and perhaps significantly more time devoted to individual targets. 

It must also be mentioned that RFI is a major culprit at $\nu_{obs}<1$~GHz, significantly restricting the redshift path and increasing the noise in each receiver band. Future OHM searches may benefit from interferometric observations (for which celestial RFI is uncorrelated from dish to dish), or from observations in more radio-quiet environments. The GMRT (India) and ASKAP (Australia) are promising instruments for interferometry, while the upcoming 64-m Sardinia Radio Telescope (Italy) and 500-m FAST (China) will be options to continue the search for OHMs at high redshifts. 

\section{Acknowledgments}

This work was part of the Ph.D. thesis of KWW at the University of Colorado, and will be submitted in early 2012 as Willett, Darling, Kent, \& Braatz (2012).  Observations were funded in part by student support grants from NRAO. We are indebted to the staff of the Green Bank Telescope, whose expertise and assistance made these observations possible. The National Radio Astronomy Observatory is a facility of the National Science Foundation operated under cooperative agreement by Associated Universities, Inc.

\end{document}